\documentclass[showpacs,twocolumn]{revtex4}%
\usepackage{amsfonts}
\usepackage{amsmath}
\usepackage{amssymb}
\usepackage{amsmath}
\usepackage{graphicx}%
\providecommand{\U}[1]{\protect\rule{.1in}{.1in}}
\begin{document}
%\linenumbers
\preprint{HEP/123-qed}
\title[Two-photon ionization of Ar]{Many-electron character of two-photon above-threshold ionization of Ar}
\author{I. D. Petrov}
\author{B. M. Lagutin}
\affiliation{Rostov State University of Transport Communications, 344038 Rostov-on-Don, Russia}
\author{V. L. Sukhorukov}
\author{N. M. Novikovskiy}
\affiliation{Institute of Physics, Southern Federal University, 344090 Rostov-on-Don, Russia}
\author{Ph.V. Demekhin}
\author{A. Knie}
\author{A. Ehresmann}
\affiliation{Institute of Physics and Center for Interdisciplinary Nanostructure Science
and Technology (CINSaT), University Kassel, D-34132, Kassel, Germany}

\begin{abstract}
The absolute generalized cross sections and angular distribution parameters of
photoelectrons for the two-photon above threshold $3p$-ionization of Ar were
calculated in the exciting photon energy range from 15.76 to 36~eV. The correlation function
technique developed earlier was extended for the case when an intermediate-state function is of a continuum-type.

We show that two-photon ionization of Ar near the $3p^{4}$ threshold
to a large extent is determined by the $(3p\dashrightarrow\varepsilon d)^{2}$
two-photon absorption via the giant resonance. This many-electron
correlation causes (i) an increase of the photoionization cross sections by
more than a factor of 3; (ii) the appearance of resonances in the exciting-photon energy range of
the doubly-excited states. The predictions are supported by a good
agreement between length and velocity results obtained after taking into
account of the higher-order perturbation theory corrections.

\end{abstract}

\maketitle

\section{Introduction}

Advancements in new intense and tunable free-electron lasers (FELs)
\cite{ackermann07s,shintake08s,emma10s,ishikawa12s,kang17s} covering the
ultra-violet and X-ray photon-energy ranges inspire a renewed
interest for a detailed comprehension of multi-photon processes.
Experimental data obtained using FELs are of great motivation for theory to
provide not only qualitative but also quantitative description of new phenomena.

One of these cases is atomic two-photon above-threshold ionization (ATI) when
the energy of a single photon is sufficient to ionize an atom. It has been shown
\cite{pindzola75,starace87,lhuillier87,pan90,petrov16,lagutin17}
that below the one-photon ionization threshold many-electron
correlations significantly influence the calculated cross sections
and angular distribution parameters of photoelectrons with respect to those
obtained in single-electron approximation. The intermediate-state shake-up
correlation studied in \cite{pan90,petrov16,lagutin17} resulted in a noticeably
closer agreement of the generalized two-photon ionization cross section (G2PICS)
calculated in length (G2PICS-$L$) and velocity (G2PICS-$V$) forms of the electric dipole
operator. The final-state electron scattering correlations taken into account
in \cite{starace87,lhuillier87,petrov16,lagutin17} decrease noticeably the absolute
values of G2PICS. In contrast, the polarization of the atomic core by the
photoelectron \cite{petrov16,lagutin17} increases the cross sections by 15-20\%.

To the best of our knowledge, there is only one \emph{ab initio} theoretical
description of the two-photon ATI of Ar \cite{pi10}. Those calculations were carried out
using Herman-Skillman \cite{herman63} potential for computing atomic orbitals (AOs)
without considering the collective behaviour of electrons.

In this work we intend to take into account many-electron correlations in the
calculation of the $3p$-ATI of Ar. To solve this problem, the correlation
function (CF) technique developed by us in \cite{petrov16,lagutin17} for the
exciting-photon energy region below the one-photon ionization threshold is applied. We extent
the CF technique for the ATI case solving two challenging problems: (i)
computing the CFs for positive energy, i.e., in the
continuum, satisfying the correct boundary conditions; (ii) computing
the free-free dipole transition matrix elements containing the CF and
final-state wave function, both of continuum-type.

The paper is organized as follows. In section \ref{sec:Amplitudes2} we
describe the correlation function method for a two-photon transition
amplitude calculation in the above-threshold exciting-photon energy region. The
technique of the matrix element calculation for the two continuum-state functions
is also described in detail. In section \ref{sec:Crosssection2} the developed
method is applied to the calculation of partial and total G2PICS of the $3p$ shell of
Ar above the $3p$ threshold. Various correlation effects are considered and
discussed. The influence of many-electron correlations on the
angular-distribution parameters of the photoelectrons is studied in section
\ref{sec:beta4}. We conclude with a brief summary in section
\ref{sec:summary5}.

\section{Two-photon transition amplitudes for the above-threshold photon
energies}

\label{sec:Amplitudes2}

In order to investigate the influence of many-electron correlations on the
ATI of atoms we considered the two-photon $3p$-ionization of
argon applying the \emph{LS}-coupling scheme:%

\begin{equation}
\mathrm{Ar}~3p^{6}(^{1}S)+2\gamma\rightarrow\mathrm{Ar}~3p^{5}(^{2}%
P)\varepsilon\ell(^{1}L)\;\;\;\;\;(L=0;2). \label{eq:scheme}%
\end{equation}

Here $L$ and $\ell$ are the orbital angular momenta of the final state and
photoelectron, respectively. We investigate the case when the energy of a
single photon is sufficient to ionize the atom. This process is known as above-threshold
ionization (ATI).

The amplitude of the two-photon ionization transition $i\rightarrow f$ with
photon energy $\omega$ in the lowest order of perturbation theory (LOPT) in
the ATI energy region is given by:%

\begin{equation}
T_{i\rightarrow f}=\sum_{m}\frac{\left\langle f\left\vert D\right\vert
m\right\rangle \left\langle m\left\vert D\right\vert i\right\rangle }%
{E_{i}+\omega-E_{m}+i\delta}, \;\;\;\;\; \delta\rightarrow0+
\label{eq:general1}%
\end{equation}
where $E_{i}$ and $E_{m}$ are energies of the initial and intermediate states
respectively, $D$ is the electric dipole operator, $\delta$ is a infinitesimal positive quantity,
and the sum contains all possible intermediate states $m$ including continuum ones. The contributions accounting
for many-electron correlations were taken into account in addition to the
LOPT amplitude (\ref{eq:general1}). We study all possible correlations
which are allowed in the next order of perturbation theory due to Coulomb
interaction of electrons. These transitions can schematically be presented as shown below.

The LOPT processes are:%

\[%
\begin{array}
[c]{lcc}%
p^{6}\dashrightarrow p^{5}\varepsilon^{\prime}\ell^{\prime}\dashrightarrow
p^{5}\varepsilon\ell &  & (\text{Ia})\\
s^{2}p^{6}\dashrightarrow s^{1}p^{6}\varepsilon\ell\dashrightarrow s^{2}%
p^{5}\varepsilon\ell &  & (\text{Ib})
\end{array}
\]
Here and below, the dash arrows denote electric dipole interaction and solid
arrows - Coulomb interaction of electrons.

The correlations described by the next order of perturbation theory are classified as follows:

\emph{Intermediate-state interchannel correlation:}%
\[%
\begin{array}
[c]{lcc}%
p^{6}\dashrightarrow p^{5}\varepsilon^{\prime\prime}\ell^{\prime\prime
}\rightarrow p^{5}\varepsilon^{\prime}\ell^{\prime}\dashrightarrow
p^{5}\varepsilon\ell &  & (\text{II})
\end{array}
\]

\emph{Ground-state correlations:}%
\[%
\begin{array}
[c]{lcc}%
p^{6}\rightarrow p^{4}\varepsilon^{\prime}\ell^{\prime}\varepsilon
^{\prime\prime}\ell^{\prime\prime}\dashrightarrow p^{5}\varepsilon^{\prime
}\ell^{\prime}\dashrightarrow p^{5}\varepsilon\ell &  & (\text{IIIa})\\
p^{6}\rightarrow p^{4}\varepsilon^{\prime}\ell^{\prime}\varepsilon
^{\prime\prime}\ell^{\prime\prime}\dashrightarrow p^{4}\varepsilon
\ell\varepsilon^{\prime\prime}\ell^{\prime\prime}\dashrightarrow
p^{5}\varepsilon\ell &  & (\text{IIIb})
\end{array}
\]%
\[%
\begin{array}
[c]{lcc}%
p^{6}\rightarrow p^{4}\varepsilon\ell\varepsilon^{\prime\prime}\ell
^{\prime\prime}\dashrightarrow p^{4}\varepsilon\ell\varepsilon^{\prime}%
\ell^{\prime}\dashrightarrow p^{5}\varepsilon\ell &  & (\text{IIIc}).
\end{array}
\]

\emph{Intermediate-state shake-up correlation:}%
\[%
\begin{array}
[c]{lcc}%
p^{6}\dashrightarrow p^{5}\varepsilon^{\prime}\ell^{\prime}\rightarrow
p^{4}\varepsilon\ell\varepsilon^{\prime}\ell^{\prime}\dashrightarrow
p^{5}\varepsilon\ell &  & (\text{IV}).
\end{array}
\]

\emph{Intermediate-state electron scattering correlations:}%
\[%
\begin{array}
[c]{lcc}%
p^{6}\dashrightarrow p^{5}\varepsilon^{\prime}\ell^{\prime}\rightarrow
p^{4}\varepsilon\ell\varepsilon^{\prime\prime}\ell^{\prime\prime
}\dashrightarrow p^{5}\varepsilon\ell &  & (\text{V}).
\end{array}
\]

\emph{Final-state electron scattering correlations:}%

\[%
\begin{array}
[c]{lcc}%
p^{6}\dashrightarrow p^{5}\varepsilon^{\prime}\ell^{\prime}\dashrightarrow
p^{4}\varepsilon^{\prime}\ell^{\prime}\varepsilon^{\prime\prime}\ell
^{\prime\prime}\rightarrow p^{5}\varepsilon\ell &  & (\text{VIa})\\
p^{6}\dashrightarrow p^{5}\varepsilon^{\prime\prime}\ell^{\prime\prime
}\dashrightarrow p^{4}\varepsilon^{\prime}\ell^{\prime}\varepsilon
^{\prime\prime}\ell^{\prime\prime}\rightarrow p^{5}\varepsilon\ell &  &
(\text{VIb}).
\end{array}
\]

\subsection{Correlation function for positive energy}

The radial part of the amplitude for process (Ia) is%

\begin{equation}
t_{\omega}^{(\text{Ia})}(L,\ell,\ell^{\prime})=\sum_{\varepsilon^{\prime}%
>F}\frac{\left\langle \varepsilon\ell\left\vert d_{r}\right\vert
\varepsilon^{\prime}\ell^{\prime}\right\rangle \left\langle \varepsilon
^{\prime}\ell^{\prime}\left\vert d_{r}\right\vert 3p\right\rangle }%
{\omega-E_{3p}^{(i)}-\varepsilon^{\prime}+i\delta}, \label{eq:radpart}%
\end{equation}
where $E_{\text{3p}}^{(i)}$ is the ionization potential of the $3p-$electron in Ar,
$d_{r}$ is the radial part of the dipole transition operator obtained in the
length or velocity form, and the notation $\varepsilon^{\prime}>F$ denotes the
summation over all unoccupied single-electron states. The wave function of the
photoelectron, $\vert\varepsilon\ell\rangle$, depends on the orbital momentum
$L$ of the final state.

The infinite summation in (\ref{eq:radpart}) can be efficiently performed by
the correlation-function (CF) method (\cite{petrov16,lagutin17} and
references therein). In the ATI case, the CF should satisfy the outgoing-wave
boundary conditions and it is a solution of the inhomogeneous
integro-differential equation%

\begin{align}
\left(  h_{\ell^{\prime}}-\omega+E_{3p}^{(i)}\right)  \phi_{\ell^{\prime}}(r)
&  =-d_{r}P_{3p}(r)\nonumber\\
&  +\sum_{n^{\prime}<F}P_{n^{\prime}\ell^{\prime}}(r)\left\langle n^{\prime
}\ell^{\prime}\left\vert d_{r}\right\vert 3p\right\rangle , \label{eq:CFunc}%
\end{align}
where $h_{\ell^{\prime}}$ is the Hartree-Fock operator for the $\varepsilon
^{\prime}\ell^{\prime}$ function in the configuration $3p^{5}\varepsilon
^{\prime}\ell^{\prime}(^{1}P)$.

In order to obtain the continuum-type CF, we have used the technique similar to that
presented in \cite{Robicheaux93}. First, the inhomogeneous equation
(\ref{eq:CFunc}) is solved disregarding the boundary conditions at large
distances ($r\rightarrow\infty$). This solution, which we
denote $\Lambda_{l^{\prime}}$, has a nonzero contribution from the incoming waves, and
its asymptotic behaviour at ($r\rightarrow\infty$) can be expressed as%

\begin{equation}
\Lambda_{l^{\prime}}(r)=gG(r)+hH(r), \label{eq:InhomoSol}%
\end{equation}
where $G(r)$ and $H(r)$ are regular and irregular Coulomb functions,
respectively \cite{curtis64}, asymptotically denoted as%

\begin{equation}
G_{\varepsilon\ell}(r)\overset{r\rightarrow\infty}{\longrightarrow}\sqrt
{\frac{2}{\pi k}}\sin\left(  kr-\frac{\ell\pi}{2}+\frac{Z}{k}\mathrm{ln}%
(2kr)+\delta_{\ell}\right) , \label{asymptG}%
\end{equation}

\begin{equation}
H_{\varepsilon\ell}(r)\overset{r\rightarrow\infty}{\longrightarrow}%
-\sqrt{\frac{2}{\pi k}}\cos\left(  kr-\frac{\ell\pi}{2}+\frac{Z}{k}%
\mathrm{ln}(2kr)+\delta_{\ell}\right)  . \label{asymptH}%
\end{equation}
Here, $k$ is the wave vector of the continuum electron in atomic units, $Z$ is
the asymptotic charge of the ionic core, $\delta_{\ell}$ represents the sum%

\begin{equation}
\delta_{\ell}=\mathrm{arg}~\Gamma\left(  \ell+1-\imath\frac{Z}{k}\right)
+\varphi_{\ell},
\end{equation}
where $\varphi_{\ell}$\ is the short-range phase shift.
Coefficients $g$ and $h$ in (\ref{eq:InhomoSol}) are defined by matching the
respective numerical solutions of inhomogeneous Eq.~(\ref{eq:CFunc}) starting
upward from zero and downward from infinity.

In order to obtain the solution of Eq.~(\ref{eq:CFunc}), which has the correct boundary
condition appropriate for the photoionization case, we add the general solution $P
_{\varepsilon^{\prime}\ell^{\prime}}(r)$ of the homogeneous equation to $\Lambda
_{l^{\prime}}(r)$, factorized with a coefficient $A$%

\begin{equation}
\left(  h_{\ell^{\prime}}-\omega+E_{3p}^{(i)}\right)  P_{\varepsilon^{\prime
}\ell^{\prime}}(r)=0. \label{eq:CFuncHomo}%
\end{equation}
The solution of Eq.~(\ref{eq:CFunc}), which has the correct boundary conditions,
is thus given by:

\begin{equation}
\phi_{l^{\prime}}(r)=\Lambda_{l^{\prime}}(r)+A\, P _{\varepsilon^{\prime}%
\ell^{\prime}}(r), \label{eq:TotalSol}%
\end{equation}
where $\Lambda_{l^{\prime}}(r)$ is the solution (\ref{eq:InhomoSol}) of
Eq.~(\ref{eq:CFunc}) with unphysical boundary conditions.

The solution of the homogeneous equation (\ref{eq:CFuncHomo}) in its asymptotics
($r\rightarrow\infty$) can be determined as%

\begin{equation}
P _{\varepsilon^{\prime}\ell^{\prime}}(r)=G(r)-K\, H(r), \label{eq:HomoSol}%
\end{equation}
where the coefficient $K$ is obtained by matching the respective numerical
solutions of Eq.~(\ref{eq:CFuncHomo}) starting upward from zero
and downward from infinity.

Substituting Eqs.~(\ref{eq:InhomoSol}) and (\ref{eq:HomoSol}) into
(\ref{eq:TotalSol}), applying the Euler expressions $\sin x=\frac{i}%
{2}(e^{-ix}-e^{ix})$ and $\cos x=\frac{1}{2}(e^{-ix}+e^{ix})$ with
$x={kr-\frac{\ell\pi}{2}+\frac{Z}{k}\mathrm{ln}(2kr)+\delta_{\ell}}$, and
equating a factor at $e^{-ix}$ to zero, we obtain the following formulae for
the coefficient $A$ in Eq.~(\ref{eq:TotalSol}):%

\begin{equation}
\mathrm{Re}\;A=-\frac{hK+g}{K^{2}+1};~~~~~~~~\mathrm{Im}\;A=\frac{gK-h}%
{K{^{2}}+1}. \label{eq:ReImA}%
\end{equation}
After the CF is determined in the form of (\ref{eq:TotalSol}), the radial part of the
transition amplitude (\ref{eq:radpart}) takes the following form:

\begin{equation}
t_{q,\omega}^{(\text{Ia})}(L,\ell,\ell^{\prime})=\left\langle \varepsilon
\ell\left\vert d_{r}\right\vert \phi_{\ell^{\prime}}\right\rangle .
\label{eq:angrad2n}%
\end{equation}

\subsection{Calculation of electric dipole transition amplitude between two
continuum-type functions}

In Eq.(\ref{eq:angrad2n}) both the CF, $\phi_{\ell^{\prime}}(r)$, and the
final-state AO, $P_{\varepsilon\ell}(r)$, are of continuum-type. In order to calculate
the dipole integral (\ref{eq:angrad2n}) between the two continuum wave
functions, a special technique was applied. We used the method described
in \cite{aymar80,gao87}. According to this method, the dipole integral (\ref{eq:angrad2n}) is
expressed as%

\begin{equation}
\left\langle \varepsilon\ell\right\vert d_{r}\left\vert \phi_{\ell^{\prime}%
}\right\rangle =\int_{0}^{r_{0}}P_{\varepsilon\ell}(r)\;d_{r}\;\phi
_{\ell^{\prime}}(r)dr+I(r_{0},\varepsilon\ell,\varepsilon^{\prime}\ell
^{\prime}), \label{eq:2integrals}%
\end{equation}
where $r_{0}$ is a sufficiently large value of $r$ and%

\begin{equation}
I(r_{0},\varepsilon\ell,\varepsilon^{\prime}\ell^{\prime})=\int_{r_{0}%
}^{\infty}u_{\varepsilon\ell}(r)\;d_{r}\;u_{\varepsilon^{\prime}\ell^{\prime}%
}(r)dr. \label{eq:Iquantity}%
\end{equation}
The radial functions $u_{\varepsilon\ell}(r)$ in (\ref{eq:Iquantity}) are the standard Coulomb functions which are the solutions of the equation%

\begin{multline}
\left(  \frac{d^{2}}{dr^{2}}+f_{\ell}(r)\right)  u_{\varepsilon\ell
}(r)=0,\label{eq:eq_u}\\
f_{\ell}(r)=k^{2}+\frac{2Z}{r}-\frac{\ell(\ell+1)}{r^{2}},~~~(r\geq r_{0}),
\end{multline}
where $\varepsilon\text{(Ry)}=k^{2}$.

The solutions of Eq.~(\ref{eq:eq_u}) are asymptotically expressed as%

\begin{equation}
u_{\varepsilon\ell}(r)=\sqrt{\frac{2}{\pi\xi_{\ell}(r)}}\sin\left(  \Phi
_{\ell}^{(1)}(r)+\delta_{\ell}\right)  . \label{eq:solution_u}%
\end{equation}
In Eq.~(\ref{eq:solution_u}), an approximate value of $\Phi_{\ell}^{(1)}(r)$
with an accuracy of $1/r^{4}$ can be obtained using the formulae presented in
\cite{aaberg82,burgess63}:%

\begin{multline}
\Phi_{\ell}^{(1)}=x+\frac{1}{m}\ln\left(  \frac{1}{m}+mp+x\right)  -\frac
{1}{m}-\frac{\ell\pi}{2}\label{eq:express_Fi1}\\
+y-\frac{x\left(  3m^{2}t+4\right)  +m\rho\left(  3m^{2}+2\right)
+mt}{24\left(  1+m^{2}t\right)  \rho}\\
+\frac{5\left(  \rho-t\right)  }{24x^{3}},
\end{multline}
where the abbreviations $m=k/Z$, $\rho=Zr$, $x=\left(  m^{2}\rho^{2}%
+2\rho-t\right)  ^{1/2}$, $t=\ell(\ell+1)$ and%

\begin{equation}
y=\left\{
\begin{array}
[c]{c}%
\frac{t+1/8}{\sqrt{t}}\arccos\left[  \frac{\rho-t+mtx}{\left(  1+m^{2}%
t\right)  \rho}\right]  ,\;\;l>0;\\
\frac{1}{4\left(  x+m\rho\right)  },\;\;\;l=0.
\end{array}
\right.  \label{eq:y}%
\end{equation}
are used. Note that in \cite{aaberg82} there is a misprint in the
definition of $x$, whereas in \cite{burgess63} there is a misprint in
Eq.~(\ref{eq:express_Fi1}).

The amplitude function $\xi_{\ell}(r)$ appearing in Eq.~(\ref{eq:solution_u}),
satisfies the following differential equation%

\begin{equation}
\xi_{\ell}^{2}(r)=f_{\ell}(r)+\xi_{\ell}^{1/2}(r)\frac{d^{2}}{dr^{2}}\xi
_{\ell}^{-1/2}(r). \label{eq:eq_xi}%
\end{equation}
Equation (\ref{eq:eq_xi}) can be solved iteratively. A zeroth approximation is
expressed as $\xi_{\ell}^{(0)}(r)=\sqrt{f_{\ell}}$. The next
approximation can be calculated by \cite{aaberg82}%

\begin{multline}
\xi_{\ell}^{(1)}=\sqrt{f_{\ell}+\xi_{\ell}^{1/4}\frac{d^{2}}{dr^{2}}\xi_{\ell
}^{-1/4}}\label{eq:eq_xi1}\\
=\sqrt{f_{\ell}+\frac{5}{16}\left(  \frac{f_{\ell}^{\prime}}{f_{\ell}}\right)
^{2}-\frac{1}{4}\frac{f_{\ell}^{\prime\prime}}{f_{\ell}}}.
\end{multline}

Integral (\ref{eq:Iquantity}) can be expressed as a difference of two terms:%

\begin{equation}
I(r_{0},\varepsilon\ell,\varepsilon^{\prime}\ell^{\prime})=I^{+}-I^{-},
\label{eq:Isum}%
\end{equation}
where%

\begin{equation}
I^{\pm}=\underset{\epsilon\rightarrow0}{\lim}\int_{r_{0}}^{\infty}e^{-\epsilon
r}\xi^{\pm}(r)g^{\pm}(r)\cos\chi^{\pm}(r)dr, \label{eq:ipm}%
\end{equation}

\begin{equation}
\xi^{\pm}(r)=\xi_{\ell^{\prime}}^{(1)}(r)\pm\xi_{\ell}^{(1)}(r),
\label{eq:xipm}%
\end{equation}

\begin{equation}
g^{\pm}(r)=d_{r}\left[  \pi\left(  \xi_{\ell^{\prime}}^{(1)}\xi_{\ell}%
^{(1)}\right)  ^{1/2}\xi^{\pm}\right]  ^{-1}, \label{eq:gpm}%
\end{equation}

\begin{equation}
\chi^{\pm}(r)=\left[  \Phi_{\ell^{\prime}}^{(1)}+\delta_{\ell^{\prime}%
}\right]  \pm\left[  \Phi_{\ell}^{(1)}+\delta_{\ell}\right]  . \label{eq:hipm}%
\end{equation}

After subsequent integration of Eq.~(\ref{eq:ipm}) by parts, the following
expansion can be obtained \cite{gao87}:%

\begin{equation}
I^{\pm}=\left\{  \sum_{n=0}^{\infty}\left[  \left(  \frac{1}{\xi^{\pm}}%
\frac{d}{dr}\right)  ^{n}g^{\pm}(r)\right]  \sin\left(  \chi^{\pm}%
(r)+\frac{n\pi}{2}\right)  \right\}  _{r=r_{0}}. \label{eq:Iplusminus}%
\end{equation}
In the present calculation we included the first four terms of expansion
(\ref{eq:Iplusminus}).

\subsection{Additional computational remarks}

The wave functions of the atomic core of Ar were computed on the ground-state
configuration $1s^{2}2s^{2}2p^{6}3s^{2}3p^{6}$ obtained in the Hartree-Fock (HF) approximation.
The final-state wave functions were obtained by solving the HF equations for a photoelectron in the
configurations $3p^{5}\varepsilon p(^{1}S), 3p^{5}\varepsilon p(^{1}D)$ and
$3p^{5}\varepsilon f(^{1}D)$ using frozen core functions. The same frozen
core functions were used in the $h_{\ell^{\prime}}$ operator in
Eq.~(\ref{eq:CFunc}).

The Hartree-Fock operator entering Eq.~(\ref{eq:CFunc}) includes the non-local
exchange part of the Coulomb interaction of the CF with core electrons. In the
same way as in \cite{petrov16}, a separate local differential equation is
determined for each term of the exchange Coulomb potential. As a result, the
single non-local integro-differential equation (\ref{eq:CFunc}), is
transformed to a system of local differential equations. In order to solve this system a
non-iterative numerical procedure is applied which is stable and converges at
each energy.

The calculation of the transition amplitude (\text{Ib}) via the $3s$ shell is
identical for above (ATI) and below threshold ionization. The latter case has
been described in our previous work \cite{petrov16}. The correlation transition amplitudes
(II)-(V) were calculated using the CF technique. The respective inhomogeneous
equations were presented and discussed in \cite{petrov16}. In the ATI energy
region, the calculations of the CFs are different in two cases. First, if the
energy denominator in the expression for the transition amplitude has no pole,
then the corresponding CF is of discrete-type, consists of a real part only, and
is calculated using technique \cite{petrov16}. Second, in the case the energy
denominator has a pole, then the CF is of continuum-type, consists of real and
imaginary parts, and is computed using the method described above.

\section{Generalized two-photon ionization cross section for above-threshold
ionization of the 3p shell of Ar}

\label{sec:Crosssection2}

In order to characterize two-photon ionization quantitatively, we use the
generalized two-photon ionization cross section (G2PICS) as defined in
\cite{starace87,pan90,petrov16}. This intrinsic quantity does not contain the exciting-photon flux
and is therefore ideally suited to characterize this process. The expression for the G2PICS is:%

\begin{equation}
\sigma_{q}(\omega)=\sum_{L,\ell}\sigma_{q}(L,\ell,\omega), \label{eq:sigma}%
\end{equation}
where $q=0$ and $q=\pm1$ correspond to a linearly or circularly polarized
incoming radiation, respectively. Each partial generalized cross section is
determined in cm$^{4}~$s and expressed via the two-photon transition
amplitudes $T_{q,\omega}(L,\ell)$ \cite{petrov16} as%

\begin{equation}
\sigma_{q}(L,\ell,\omega)=\frac{8\pi^{3}\alpha a_{0}^{5}}{c}\omega^{\pm
2}\left\vert T_{q,\omega}(L,\ell)\right\vert ^{2}. \label{eq:sigmapart}%
\end{equation}
In Eq.~(\ref{eq:sigmapart}), $\alpha=1/137.036$ is the fine-structure constant;
$a_{0}=5.29177\cdot10^{-9}~$cm is the Bohr radius; $c=2.99792\cdot10^{10}%
$~cm/s is the light velocity in vacuum; $\omega$ is the exciting photon energy in
atomic units; `$+$' and `$-$' correspond to the length and velocity form of
the electric dipole transition operator. The major details of the transition amplitude
calculation (particularly the calculation of the angular parts) has been described in
our previous work \cite{petrov16}.

\subsection{LOPT approximation}

\begin{figure}[ptbh]
\begin{center}
\includegraphics[width=0.47\textwidth]{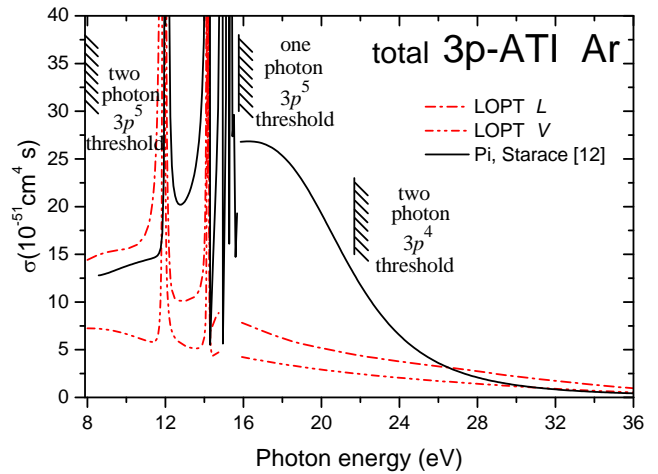}
\end{center}
\caption{The 3$p$ shell total G2PICS in length ($L$) and velocity ($V$) form
for linearly polarized incoming radiation calculated in LOPT below the
one-photon 3$p$ threshold \cite{petrov16} and in the ATI region (present work).
The calculation of \cite{pi10} is also presented for comparison (see text).
The two-photon $3p^{5}(^{2} P_{3/2})$ and $3p^{4}(^{3} P_{2})$
thresholds from \cite{kramida14} are depicted as hatched lines.}%
\label{fig:01}%
\end{figure}

The G2PICS of the $3p$ shell of Ar calculated in LOPT
approximation (processes (Ia) and (Ib)) are presented in Fig.~\ref{fig:01} in
the exciting-photon energy region below the one-photon ionization threshold (our
calculation from \cite{petrov16}) and in ATI region (present calculation). In
the same figure, the results of the G2PICS of Ar computed in \cite{pi10} in the LOPT approach
is also depicted.

The total G2PICS have been plotted in \cite{pi10} vs the energy $\varepsilon
_{1}=-\left\vert \varepsilon_{3p}\right\vert +\omega$ of the electron in the
intermediate state. In our calculation the G2PICS is presented as functions of the
exciting-photon energy $\omega$. To compare our data with the data of \cite{pi10}
we used the experimental one-photon ionization potential of the
$3p$-electron $E_{3p}^{(i)}=1.158$~Ry (corresponding to the energy level
$E(^{2} P_{3/2})=15.7596$~eV \cite{kramida14}) instead of the Herman-Skillman value
$\left\vert \varepsilon_{3p}\right\vert =1.065$~Ry used in \cite{pi10}. This
aligns the one-photon ionization thresholds in both calculations (see hatched
line at 15.76~eV in Fig.~\ref{fig:01}).

The G2PICS obtained in \cite{pi10} agree fairly well with our calculation close to
the two-photon ionization threshold ($\omega=7.8798$~eV). The data of
\cite{pi10} is located between our G2PICS-$L$ and G2PICS-$V$ and differs from our
G2PICS-$L$ by 15\%. The deviation is larger in the ATI region, as seen from
Fig.~\ref{fig:01}. At the one-photon $3p^{5}$ threshold, the cross section from
\cite{pi10} exceeds our G2PICS-$L$ by more than three times,
whereas at high photon energies it decreases more rapidly than our G2PICS, after
$\omega=30$~eV it is close to the present G2PICS-$V$. We suggest that this disagreement arises most likely from
the different approximations in the CFs calculation: Herman-Skillman in \cite{pi10} vs term-dependent
Hartree-Fock 3$p^{5} \phi_{\ell^{\prime}}(^{1}\text{P})$ in the
present work.

In order to support this suggestion we performed a separate study by calculating the G2PICS
in the LOPT approach using the average self-consistent
$3p^{5}\varepsilon^{\prime}\ell^{\prime}$ configuration instead of the
term-dependent frozen core $3p^{5}\varepsilon^{\prime}\ell^{\prime} (^{1}P)$
configuration. In this case the present cross sections became closer to those
obtained in \cite{pi10}.

The value of amplitude (Ib) for the transition via the $3s$ shell is less than
0.1\% of the amplitude (Ia) and has practically no influence on the calculated
cross section.

The results depicted in Fig.~\ref{fig:01} show that the G2PICS
calculated in our work in LOPT approximation in the length and velocity forms
differ from each other by approximately two times. This difference indicates the
necessity to include many-electron correlation in the calculation.

\subsection{Correlation processes of third order of perturbation theory}

The total two-photon ionization cross section of the $3p$ shell of Ar is a sum
of three partial cross sections for the transitions to the $\varepsilon
p(^{1}S)$, $\varepsilon p(^{1}D)$ and $\varepsilon f(^{1}D)$ channels. As in
\cite{pi10}, the $\varepsilon f(^{1}D)$ channel is found to dominate in the
ATI region: in LOPT approximation at the one-photon $3p^{5}$ threshold the partial
$\varepsilon f(^{1}D)$ cross sections are about 85\% and 89\% of the total 3$p$ cross
sections in the length and velocity gauge, respectively. Therefore, the
influence of many-electron correlations will be demonstrated here in detail for the
case of the $\varepsilon f(^{1}D)$ channel only.

\begin{figure}[ptbh]
\begin{center}
\includegraphics[width=0.47\textwidth]{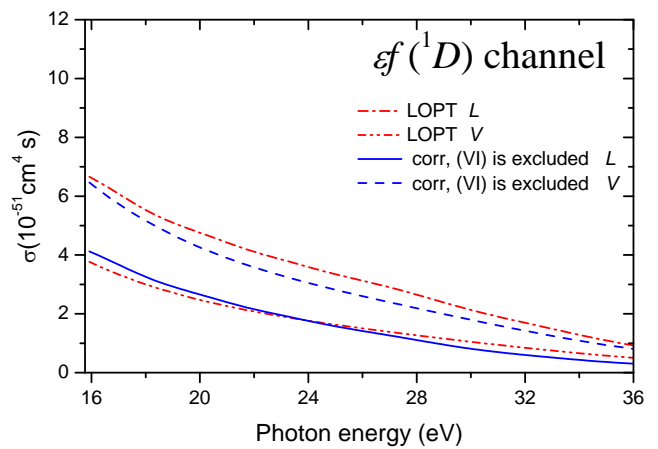}
\end{center}
\caption{Partial G2PICS for the transition to the $3p^{5}\varepsilon f(^{1}D)$
channel, computed in length ($L$) and velocity ($V$) form for linearly
polarized incoming radiation in LOPT and with accounting for transitions
(I)-(V) (corr (VI) is excluded).}%
\label{fig:02}%
\end{figure}

In Fig.~\ref{fig:02}, the partial G2PICS for the $\varepsilon f(^{1}D)$
channel calculated in LOPT are compared with the cross section computed with
accounting for the processes (I)--(V) (correlation (VI) is excluded). The
starting value of the photon energy corresponds to the one-photon $3p^{5}$-ionization
threshold $\omega=15.7596$~eV. In Fig.~\ref{fig:02}, it is seen that the
correlative transitions (II)-(V) change the G2PICS only quantitatively but
without any qualitative change in their energy dependence.

It turned out that in $3p$-ATI correlation correction (\text{VI}) is
much more important than below the one-photon $3p$ threshold. In ATI, the transition amplitude
(\text{VI}) is substantially (one order of magnitude) larger than the other
correlation amplitudes (II)--(V) and changes the computed cross sections drastically.
In the processes (\text{VIa}) and (\text{VIb}), the first and the second
photon excite a pair of $3p$ core electrons to virtual $\varepsilon
^{\prime}\ell^{\prime}$ and $\varepsilon^{\prime\prime}\ell^{\prime\prime}$
states; then because of the Coulomb interaction, one electron returns to the
core and another changes the state to the $\varepsilon\ell$ final one. The large
value of this correlation can be explained through the existence of the giant resonance in each
of the single-electron $\left\langle \varepsilon d\left\vert d_{r}\right\vert
3p\right\rangle $ transition amplitudes.

The radial part of the transition amplitude (VI) is described by the following expression:%

\begin{multline}
t_{\omega}^{(\text{VIa,b})}(L,\ell,\ell^{\prime},\ell^{\prime\prime})=\\
=\sum_{\varepsilon^{\prime},\varepsilon^{\prime\prime}>F}\underset{k}{\sum
}\left[  a_{k}R^{k}(\varepsilon\ell3p,\varepsilon^{\prime}\ell^{\prime
}\varepsilon^{\prime\prime}\ell^{\prime\prime})\right. \\
\left.  +b_{k}R^{k}(\varepsilon\ell3p,\varepsilon^{\prime\prime}\ell
^{\prime\prime}\varepsilon^{\prime}\ell^{\prime})\right]  \left\langle
\varepsilon^{\prime\prime}\ell^{\prime\prime}\left\vert d_{r}\right\vert
3p\right\rangle \left\langle \varepsilon^{\prime}\ell^{\prime}\left\vert
d_{r}\right\vert 3p\right\rangle \\
\times\left\{  \frac{1}{(2\omega-E_{3p^{2}}^{(i)}-\varepsilon^{\prime
}-\varepsilon^{\prime\prime})(\omega-E_{3p}^{(i)}-\varepsilon^{\prime}%
)}\right. \label{eq:screen14}\\
\left.  +\frac{1}{(2\omega-E_{3p^{2}}^{(i)}-\varepsilon^{\prime}%
-\varepsilon^{\prime\prime})(\omega-E_{3p}^{(i)}-\varepsilon^{\prime\prime}%
)}\right\}  ,
\end{multline}
where $E_{3p^{2}}^{(i)}$ $=3.189$~Ry is the experimental $43.3893$~eV \cite{kramida14} energy of the
$3p^{4}(^{3}P_{2})$ level. The $a_{k}$ and $b_{k}$ are
numerical coefficients given in \cite{petrov16}, and $R^{k}(\varepsilon
\ell3p,\varepsilon^{\prime}\ell^{\prime}\varepsilon^{\prime\prime}\ell
^{\prime\prime})$ is the Slater integral%

\begin{multline}
R^{k}(n_{1}\ell_{1},n_{3}\ell_{3};n_{2}\ell_{2},n_{4}\ell_{4})=\\
=\int_{0}^{\infty}P_{n_{3}\ell_{3}}(r)P_{n_{4}\ell_{4}}(r)y_{k}(P_{n_{1}%
\ell_{1}},P_{n_{2}\ell_{2}};r)dr, \label{Slater}%
\end{multline}

\begin{equation}
y_{k}(P_{n_{1}\ell_{1}},P_{n_{2}\ell_{2}};r) =\int_{0}^{\infty}\frac{r_{<}%
^{k}}{r_{>}^{k+1}}P_{n_{1}\ell_{1}}(r^{\prime})P_{n_{2}\ell_{2}}(r^{\prime
})dr^{\prime}, \label{eq:jk}%
\end{equation}
where $r_{<}$ and $r_{>}$ is the smaller and the larger of the radial coordinates $r$ and
$r^{\prime}$.

Since the first factor in the denominator of the two terms in the curly
braces of Eq.~(\ref{eq:screen14}) contains both the $\varepsilon^{\prime}$ and
$\varepsilon^{\prime\prime}$ intermediate state energies, the amplitude
(\ref{eq:screen14}) cannot be expressed via CFs.
In \cite{petrov16}, we followed the approximation of
\cite{pan90}. In the first factor of the first fraction in curly braces
containing $2\omega$, we replaced $\varepsilon^{\prime}$ by $\varepsilon
^{\prime\prime}$ whereas in the second fraction in curly braces, we replaced
$\varepsilon^{\prime\prime}$ by $\varepsilon^{\prime}$. After this assumption,
Eq.~(\ref{eq:screen14}) was simplified and expressed via the correlation
functions. We estimated that at the near-threshold two-photon $3p$-ionization
region the error of the approximated amplitude did not exceed 7\% with respect
to the data obtained using the exact expression (\ref{eq:screen14}).

The electron-scattering correlation (\text{VI}) is found to be very important
in the ATI region. Therefore, we calculated it without the approximation discussed
above using the exact expression (\ref{eq:screen14}). This was
possible in this special case because in Eq.~(\ref{eq:screen14}) both the Slater integrals
$R^{k}(\varepsilon\ell3p,\varepsilon^{\prime\prime}\ell^{\prime\prime
}\varepsilon^{\prime}\ell^{\prime})$ and the dipole integrals
$\left\langle\varepsilon^{\prime}\ell^{\prime}\right\vert d_{r}\left\vert 3p\right\rangle $
and
$\left\langle\varepsilon^{\prime\prime}\ell^{\prime\prime}\right\vert d_{r}\left\vert 3p\right\rangle $
are not divergent due to the presence of a localized $3p$ function. To perform the
summation in Eq.~(\ref{eq:screen14}) over the $\varepsilon^{\prime}%
\ell^{\prime}$ and $\varepsilon^{\prime\prime}\ell^{\prime\prime}$, several
discrete states and the continuum states up to 20 Ry were taken into account.

\begin{figure}[ptbh]
\begin{center}
\includegraphics[width=0.47\textwidth]{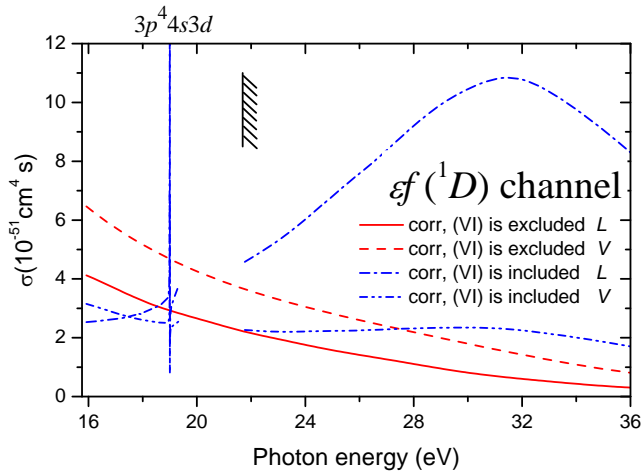}
\end{center}
\caption{Partial G2PICS for the transition to the $3p^{5}\varepsilon f(^{1}D)$
channel, computed in length ($L$) and velocity ($V$) form for linearly
polarized incoming radiation with accounting for transitions (I)-(V) (corr,
(VI) is excluded) and with accounting for all correlations (I)-(VI) (corr,
(VI) is included). The two-photon $3p^{4}(^{3}P_{2})$ ionization threshold is
indicated by the hatched line. We show the energetically lowest calculated $3p^{4}4s3d$
doubly-excited state resonance only.}%
\label{fig:03}%
\end{figure}

In Fig.~\ref{fig:03}, partial GSPICS for the transition to the $3p^{5}%
\varepsilon f(^{1}D)$ channel calculated considering the correlations
(II)-(V) only and with all correlations (II)-(VI) are compared. A drastic difference
between the two results is evident: (i) the G2PICS in
length form is enhanced by an order of magnitude resulting in a pronounced
maximum at $\omega=32$ eV; (ii) the correlation (VI) gives rise to
doubly-excited state resonances below the two-photon $3p^{4}$
double-ionization threshold. The resonances become apparent at those photon
energies when the denominators in Eq.~(\ref{eq:screen14})
containing $2\omega$ are vanishing. The first, energetically lowest
3$p^{4}3d4s$ resonance in the partial $3p^{5}\varepsilon f(^{1}D)$ G2PICS is
shown in Fig.~\ref{fig:03}. We restricted our presentation to this resonance because
the precise calculation of the doubly-excited state energies is a
cumbersome problem on its own (see, e.g.,
\cite{sukhorukov92,sukhorukov94a,kammer06}). The energies of the respective
doubly-excited state resonances are expected to be larger than they appear
in the relaxed $3p^{4}n^{\prime}\ell^{\prime}n^{\prime\prime}\ell
^{\prime\prime}$ configuration. The two-photon double-ionization threshold
equals to $0.5E_{3p^{2}}^{(i)}$ is depicted in Fig.~\ref{fig:03} by the hatched
line. It indicates the upper limit of the doubly-excited state resonances.

The revealed strong influence of many electron correlations on G2PICS
including the resonance structure is beyond a simple single-electron picture
of two-photon ionization and its experimental verification is of great fundamental importance.

\subsection{Many-electron correlations of higher orders of perturbation theory}

G2PICS calculated in length and velocity gauges considering all
the transitions (I)-(VI) still differ substantially, as it is seen from
Fig.~\ref{fig:03}. At $\omega=32$ eV the G2PICS-$L$ (dashed-dotted line) is
four times larger than G2PICS-$V$ (dash-double-dotted line). In our work,
\cite{petrov16} a good agreement between computed G2PICS-$L$ and G2PICS-$V$
was achieved after inclusions of higher-order PT correlations. Those effects
were: (i) the polarization of the atomic core by the photoelectron and (ii)
the correlational decrease of the Coulomb interaction in the description of the
correlative processes  (II)-(VI). Additional calculations showed that in ATI
region taking into account these higher-order PT correlations (i)-(ii) is not sufficient
to make an agreement between G2PICS-$L$ and G2PICS-$V$ close. The reason is the anomalously
large contribution of the correlation (VI) discussed above. In the present work, we take
into account intrashell correlations \cite{amusia75} in addition to
correlations (i)-(ii) when computing the matrix elements $\left\langle \varepsilon
^{\prime}\ell^{\prime}\right\vert d_{r}\left\vert 3p\right\rangle $ and
$\left\langle \varepsilon^{\prime\prime}\ell^{\prime\prime}\right\vert
d_{r}\left\vert 3p\right\rangle $.

The \emph{ab initio} core polarization potential $V^{CP}(r)$ \cite{petrov99} was
included in the HF operator $h_{\ell^{\prime}}(r)$ entering the equations for the
CFs and for the final-state electrons. In addition, the matrix elements of the
Coulomb interaction describing the correlations (II)-(VI) have been reduced by
a factor of 1.25 \cite{petrov16}. Here and below, the calculation with taking
into account the processes (I)-(VI), intrashell correlations, polarization of
the atomic core by the photoelectron and the correlational decrease of the Coulomb
interaction will be designated as the configuration interaction Hartree-Fock approach
with core polarization (CIHFCP).

\begin{figure}[ptbh]
\begin{center}
\includegraphics[width=0.47\textwidth]{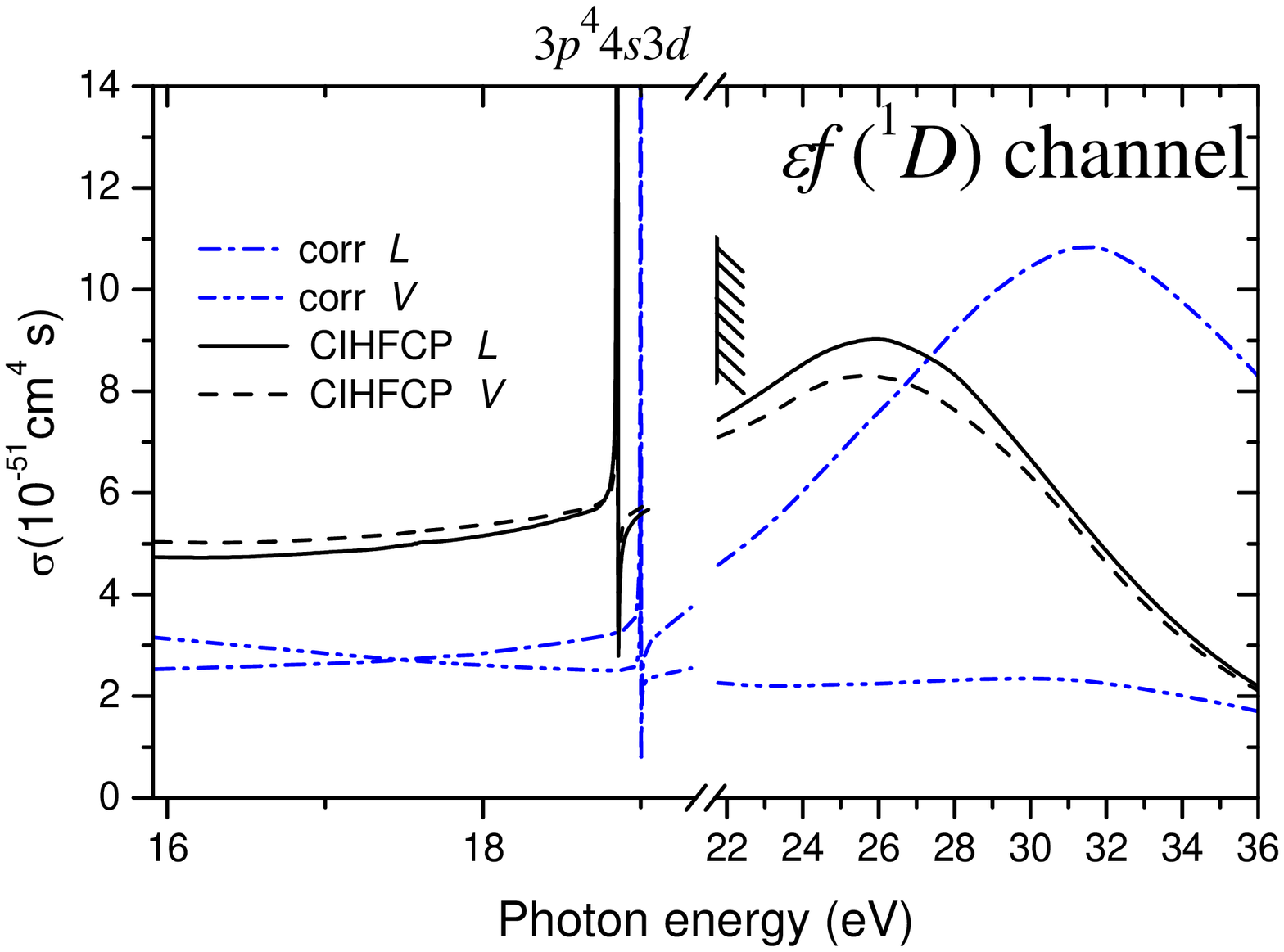}
\end{center}
\caption{Partial G2PICS for the transition to the $3p^{5}\varepsilon f(^{1}D)$
channel, computed in length ($L$) and velocity ($V$) form for linearly
polarized incoming radiation with accounting for all the correlations (I)-(VI)
(corr) and, in addition, the higher-order PT correlations (CIHFCP). The
two-photon $3p^{4}(^{3}P_{2})$ ionization threshold is indicated by the hatched
line. The first calculated $3p^{4}4s3d$ doubly-excited state resonance is shown
only.}%
\label{fig:04}%
\end{figure}

The CIHFCP partial cross sections for the $\varepsilon f (^{1}D)$ channel are
depicted in Fig.~\ref{fig:04} (solid and dashed curves for length and velocity
form, respectively). In the same figure the cross sections calculated with
taking into account processes (I)-(VI) but without higher-order PT
corrections are also presented (dash-dotted and dash-double-dotted curves) for
comparison. It is clearly seen that accounting for higher-order PT
corrections results in a very close agreement between the G2PICS-$L$ and
G2PICS-$V$ both above the two-photon double-ionization threshold (to the
right of the hatched line in Fig.~\ref{fig:04}) and in the region of the
$3p^{4}4s3d$ resonance. The following changes in the calculated G2PICS can be seen from
Fig.~\ref{fig:04}:

(i) the energy of the $3p^{4}4s3d$ resonance shifted by 0.15 eV to the
low-energy side. The reason of the shift is the polarization of the atomic core by
the photoelectron which decreases energies of the excited electrons;

(ii) the G2PICS becomes larger on the one-photon threshold ($\omega=15.76$ eV)
by about 2 times;

(iii) maximum in the $\sigma(\omega)$ dependence above the two-photon $3p^{4}$
threshold is now shifted by $\sim6$~eV to the low-energy side.

\begin{figure}[ptbh]
\begin{center}
\includegraphics[width=0.47\textwidth]{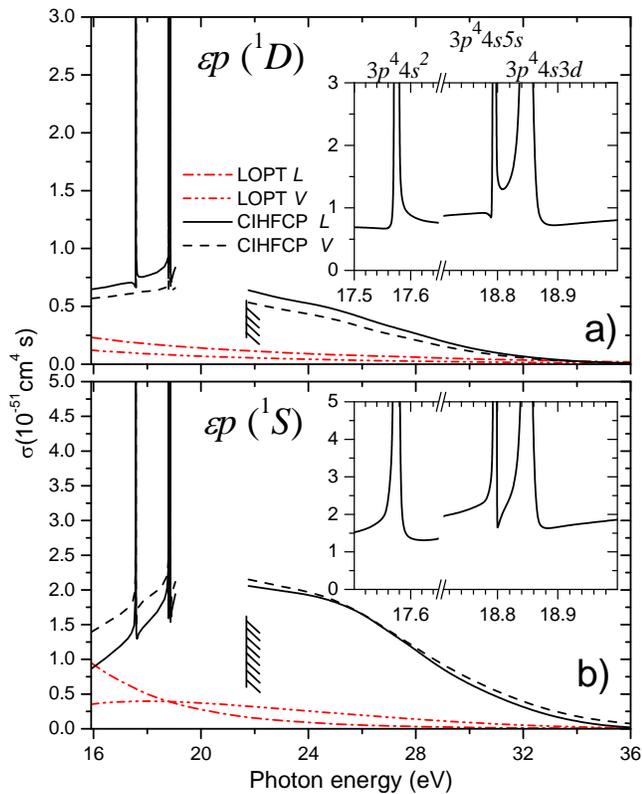}
\end{center}
\caption{Partial G2PICS for the transitions to the $3p^{5}\varepsilon p(^{1}D)$
channel (a) and to the $3p^{5}\varepsilon p(^{1}S)$ channel (b), computed in
length ($L$) and velocity ($V$) form for linearly polarized incoming
radiation in LOPT and with taking into account both the correlations (I)-(VI)
and higher-order PT correlations (CIHFCP). The two-photon $3p^{4}(^{3}P_{2})$
ionization threshold is indicated by the hatched line. }%
\label{fig:05}%
\end{figure}

The computed partial G2PICSs for the transitions to the $3p^{5}\varepsilon
p(^{1}D)$ and $3p^{5}\varepsilon p(^{1}S)$ channels are depicted in
Figs.~\ref{fig:05}a and \ref{fig:05}b, respectively. Dash-dotted and
dash-double-dotted lines represent results computed in the LOPT and solid and
dashed lines represent CIHFCP results. Similar to the $3p^{5}\varepsilon
f(^{1}D)$ channel, one can recognize a strong increase the G2PICSs, the appearance of
the doubly-excited state resonances and a better agreement between length an velocity form results.

In the insets of Figs.~\ref{fig:05}a,b the doubly-excited state resonance
region of the G2PICS-$L$ is presented on an enlarged scale. For the
$3p^{5}\varepsilon f(^{1}D)$ channel, the lowest resonance corresponds to the
$3p^{4}4s3d$ state (see Figs.\ref{fig:03} and \ref{fig:04}), whereas in the
$3p^{5}\varepsilon p(^{1}D)$ and $3p^{5}\varepsilon p(^{1}S)$ channels the
$3p^{4}4s^{2}$ and $3p^{4}4s5s$ states are also present. As it was already
mentioned, precise calculation of the resonance energies is a separate
cumbersome problem. In more detail, in the present calculation the energy of
the $4s$ electron is equal to $\varepsilon_{4s}= - 0.303$ Ry. When using the
experimental value of the ionization potential $E_{3p^{2}}^{(i)}=3.189$ Ry,
the first terms in both denominators in Eq.~(\ref{eq:screen14}) are vanishing
at $\omega=0.5\,E_{3p^{2}}^{(i)}+\varepsilon_{4s}=17.57$ eV. This
energy corresponds to the position of the $3p^{4}4s^{2}$ resonance in
Fig.~\ref{fig:05}. The experimental energy of the $3p^{4}4s^{2}$ resonance is
at $\omega=13.475$ eV \cite{jorgensen78}. The main reason for this discrepancy
is the approximate (frozen-core) value of $\varepsilon_{4s}$.

\begin{figure}[ptbh]
\begin{center}
\includegraphics[width=0.47\textwidth]{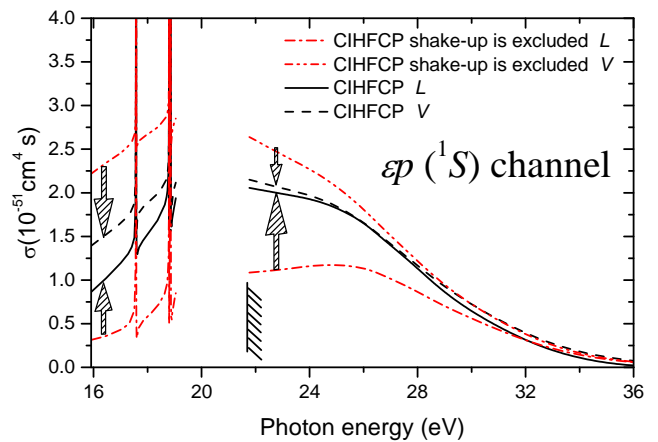}
\end{center}
\caption{Partial G2PICS for the transition to the $3p^{5}\varepsilon p(^{1}S)$
channel, computed in length ($L$) and velocity ($V$) form for linearly
polarized incoming radiation without (CIHFCP shake-up excluded) and with
(CIHFCP) taking into account the shake-up correlation (IV). The two-photon
$3p^{4}(^{3}P_{2})$ ionization threshold is indicated by the hatched line. }%
\label{fig:06}%
\end{figure}

In \cite{pan90,petrov16,lagutin17}, a large
influence of the intermediate-state shake-up (IV) correlation on the computed
G2PICS was revealed, particularly on the $3p^{5}\varepsilon p(^{1}S)$ partial cross
section. In the ATI case, this influence is of similar size, as demonstrated in
Fig.~\ref{fig:06}. Correlation (IV) influences both the real and the imaginary
part of the transition amplitude and pulls together the $3p$ G2PICS computed
in the length and velocity gauges.

\begin{figure}[ptbh]
\begin{center}
\includegraphics[width=0.47\textwidth]{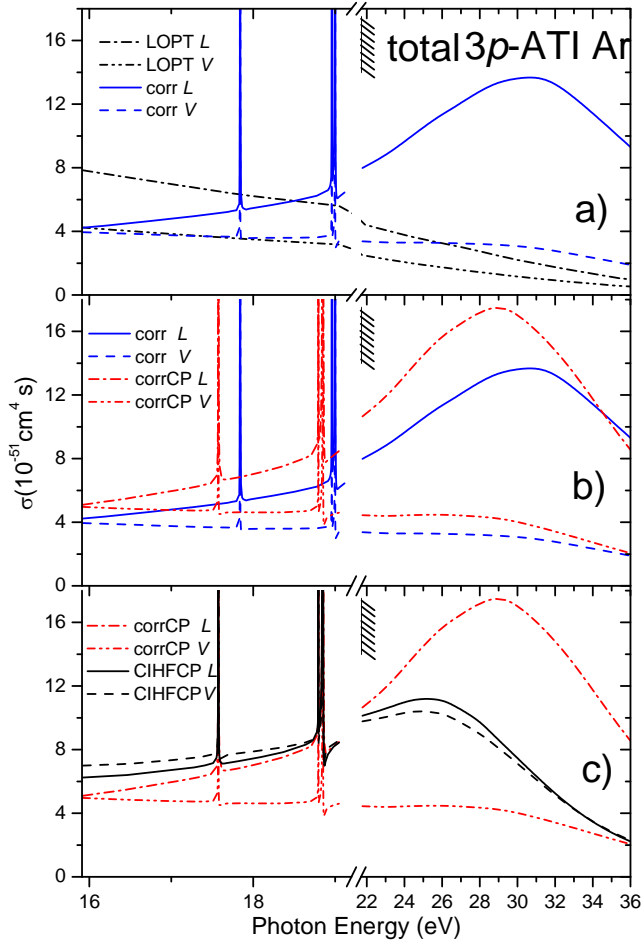}
\end{center}
\caption{The computed total $3p$ shell G2PICS in length ($L$) and velocity
($V$) form for linearly polarized incoming radiation. (a) LOPT approach
and considering all the correlations (I)-(VI) (corr); (b)
polarization of the atomic core by the photoelectron (corrCP) is included in
addition to a); (c) higher-order PT correlations (CIHFCP) are included in
addition to b). The two-photon $3p^{4}(^{3}P_{2})$ ionization threshold is
indicated by the hatched line. }%
\label{fig:07}%
\end{figure}

Concluding this section we demonstrate the effect of a successive inclusion of different
correlations in the calculated total $3p$ shell G2PICS in Fig.~\ref{fig:07}.
In Fig.~\ref{fig:07}a, we compare the G2PICS computed in LOPT approximation
with the cross sections obtained with taking into account correlations
(II-VI). The drastic change of $\sigma(\omega)$ is obvious: a
resonance structure appears below the two-photon $3p^{4}$ threshold and a
considerable enhancement of the above-threshold cross section together with
a change of the shape of $\sigma(\omega)$ from monotonic decrease to curve with a broad maximum occurs.
Those changes are mainly due to the electron-scattering correlation (VI), i.e a $(3p\dashrightarrow
\varepsilon d)^{2}$ absorption of two exciting photons at the giant resonance
with a subsequent Auger type interaction $\varepsilon d \varepsilon d - 3p
\varepsilon f$.
This mechanism should also influence the two-photon ATI of Xe in the range of its giant $4d \dashrightarrow
\varepsilon f$ resonance. An experimental proof by determining the two-photon G2PICS in Ar or Xe close
to their respective $3p^{4}$ and $4d^{8}$ thresholds would be highly desirable.

In Fig.~\ref{fig:07}b, the influence of the polarization of the atomic core by
the photoelectron on the computed total cross section is presented. It results
in a shift of the maximum of $\sigma(\omega)$ to lower photon energies and makes the
humped curve more vivid. At the two-photon $3p^{4}(^{3}P_{2})$ threshold the
G2PICS increases by $\sim 33\%$.

In Fig.~\ref{fig:07}c, the influence of the intrashell correlation on the dominating
process (VI) and the effect of the correlational decrease of the Coulomb interaction on the
transitions (II)-(VI) is shown. These higher-order PT corrections bring
the cross sections computed in length and velocity form in the considered
photon-energy regions together.

\section{Angular distribution of photoelectrons}

\label{sec:beta4}

The expression for the differential G2PICS is as follows%

\begin{multline}
\frac{d\sigma_{q}(\omega)}{d\Omega}=\frac{\sigma_{q}(\omega)}{4\pi}\\
\times\left[  1+\beta_{2}^{q}(\omega)P_{2}(\cos\theta)+\beta_{4}^{q}%
(\omega)P_{4}(\cos\theta)\right]  , \label{eq:dsdw}%
\end{multline}
where $\beta_{\lambda}^{q}$ are angular distribution parameters for
photoelectrons, $P_{\lambda}$ are Legendre polynomials, $\theta$ is the angle between
the momentum of the photoelectron and the electric field vectors for linearly
polarized incident radiation ($q$ = 0) or between the momentum of
photoelectron and the direction of propagation vectors of circularly
polarized incoming radiation ($q=\pm1$). The photoelectron angular distribution parameters $\beta_{\lambda}^{q}$ are expressed via the transition amplitudes $T_{q,\omega}(L,\ell)$ discussed above. The corresponding expressions and numerical coefficients are
reported in \cite{petrov16}.

\begin{figure}[ptbh]
\begin{center}
\includegraphics[width=0.47\textwidth]{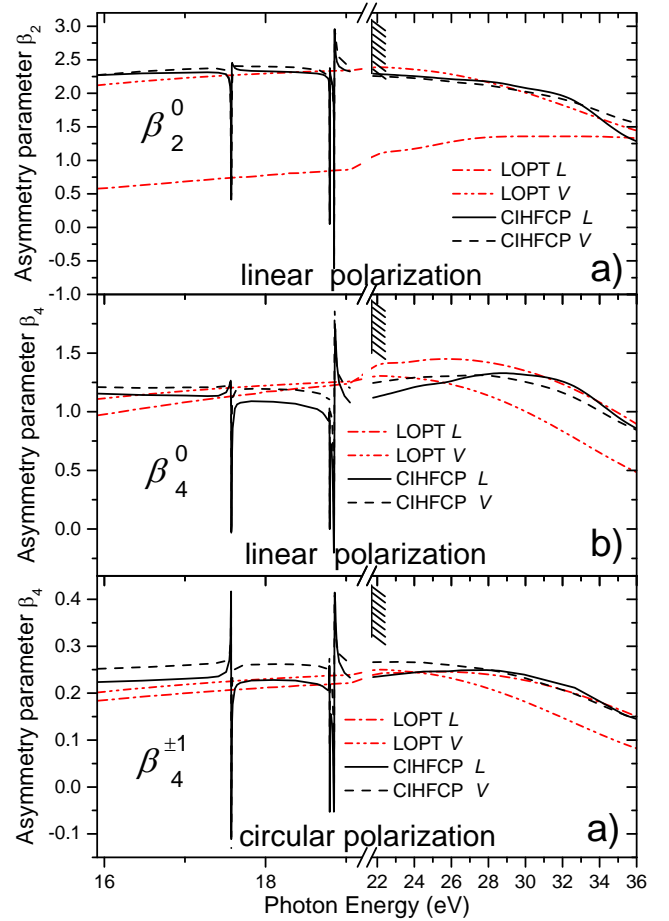}
\end{center}
\caption{ Angular-distribution parameters $\beta^{q}_{\lambda}$ for
photoelectrons computed for the two-photon 3$p$ ATI of Ar in length ($L$)
and velocity ($V$) form in LOPT approximation and with taking into account all
the considered correlations (CIHFCP). The two-photon $3p^{4}(^{3}P_{2})$
ionization threshold is indicated by the hatched line.}%
\label{fig:08}%
\end{figure}

The results of the present calculation are depicted in Figs.~\ref{fig:08}a,b,c for the
$\beta_{2}^{0}$, $\beta_{4}^{0}$, and $\beta_{4}^{\pm1}$ parameters,
respectively. The $\beta_{2}^{0}$ and $\beta_{4}^{0}$ parameters describe the
case of linearly polarized incoming radiation and $\beta_{4}^{\pm1}$
is related to circularly polarized photons. The $\beta_{2}^{\pm1}$
parameter is not presented in Fig.~\ref{fig:08} because it is determined by the connection
with $\beta_{4}^{\pm1}$ via $\beta_{2}^{\pm1}=-1-\beta_{4}^{\pm1}$ \cite{petrov16}.
In all cases taking into account many-electron correlations improves the agreement
between length and velocity results.

The comparison between $\beta(\omega)$ computed in the LOPT approximation and with
taking into account all the considered correlations exhibits a considerable
change in $\beta(\omega)$ dependencies in the range where the doubly-excited
resonances are situated. The influence of many-electron correlations on
$\beta(\omega)$ at $\omega\gtrsim22$\ eV is less pronounced than in the G2PICS
case. To a large extent this fact is connected with the prevalence of the
$3p^{5}\varepsilon f$ channel in the two-photon photoionization of Ar due to
the influence of the $3p\dashrightarrow\varepsilon d$\ giant resonance.

\section{Conclusions}

\label{sec:summary5}

In the present work, the two-photon ionization of the 3$p$ shell of Ar in the
above-threshold ionization (ATI) region was studied theoretically. For this purpose, we
extended a  noniterative correlation function (CF) technique developed earlier \cite{petrov16}
to the case when both the CF and the final-state
function are of continuum-type. An adequate accuracy of present calculations supported by good agreement between the results
obtained in length and velocity form of dipole transition operator.

The many-electron correlation which can be treated as the $(3p\dashrightarrow
\varepsilon d)^{2}$ absorption of the two exciting photons at giant resonance play a
decisive role in two-photon ionization of Ar near the $3p^{4}$ threshold.
Taking into account this correlation results in (i) an increase of the computed
G2PICS by approximately three times and (ii) the appearance of resonance
profiles in the energy range of the doubly-excited $3p^{4}n^{\prime}%
\ell^{\prime}n^{\prime\prime}\ell^{\prime\prime}$ states ($15.76<\omega<21.69$ eV).

Taking into account all remaining single and double electron excitations in
computing the transition amplitudes results in better agreement between
G2PICS-$L$ and G2PICS-$V$. Nevertheless, a 3 time difference between them
in some partial photoionization channels still remains. This difference was removed after
taking into account higher-order PT corrections. These corrections were
included by: (i) the polarization of the atomic core by the
photoelectron, realized by incorporation of an \emph{ab initio} polarization
potential in the Hamiltonian; (ii) the effective correlational decrease of the
Coulomb interaction, taken into account by computing respective
reduction coefficients; (iii) the intrashell correlations, taken into account via
a technique described in \cite{amusia75}. Good agreement between length and
velocity results for angular distribution parameters was also obtained after
inclusion of the described above many-electron correlations.

Finally, our \emph{ab initio} computation clarified that the two-photon
above-threshold $3p$ ionization of Ar near the $3p^{4}$ threshold is almost
entirely a collective-electron process. An experiment to benchmark the present
calculation is desirable.

\begin{acknowledgments}
This work was supported by the Sonderforschungsbereich SFB-1319 ``Extreme light for
sensing and driving molecular chirality'' of the Deutsche Forschungsgemeinschaft (DFG). IDP, BML and VLS would like to thank the Institute of Physics, University of Kassel for the hospitality. VLS appreciate support from Southern Federal University within the inner project no~3.6105.2017/BP. IDP and BML appreciate support from the grant
no~16-02-00666A of the RFBR.

We are grateful to Prof. A.Starace and Dr. L.-W. Pi for providing data from
fig.~2 of Ref.~\cite{pi10} in digital form. We are also grateful to Dr. A.Kramida
providing us a reference with the experimental energy of the $3p^{4}4s^{2}$
doubly-excited state of argon.
\end{acknowledgments}

%\bibliography{Petrov2015c}%
%\bibliographystyle{phpf}
%\bibliography{Library_VLS}
%\bibliography{multiphoton,mybooks,ours,dpes,doimultiphoton_august15_kl,multiphoton_VLS_ATI}

\end{document}